\documentclass[M]{iucrjournals}

\newcommand{\revision}[1]{{\textcolor{black}{#1}}}
\usepackage{siunitx}
\usepackage{gensymb}
\usepackage{amsmath}
\usepackage[utf8]{inputenc}   
\usepackage[T1]{fontenc}

\title{Probing laser-driven surface and subsurface dynamics via grazing-incidence XFEL scattering and diffraction}

\author[a,$\dagger$]{Lisa Randolph\IUCrCemaillink{Lisa.Randolph@uni-siegen.de}\IUCrOrcidlink{0000-0001-9587-404X}}%
\author[b]{\"{O}zg\"{u}l \"{O}zt\"{u}rk\IUCrOrcidlink{0000-0002-6283-4161}}%
\author[b]{Dmitriy Ksenzov\IUCrOrcidlink{0000-0002-0657-1678}}%
\author[c]{Lingen Huang\IUCrOrcidlink{0000-0003-1184-2097}}%
\author[c]{Thomas Kluge\IUCrOrcidlink{0000-0003-4861-5584}}%
\author[a]{S.V. Rahul\IUCrOrcidlink{0000-0002-8215-2264}}%
\author[a]{Victorien Bouffetier\IUCrOrcidlink{0000-0001-6079-1260}}%
\author[d]{Tobias Held\IUCrOrcidlink{0009-0009-8925-1810}}%
\author[d]{Sebastian T. Weber\IUCrOrcidlink{0000-0002-9090-2248}}%
\author[c]{Carsten Baehtz}%
\author[a,e]{Mohammadreza Banjafar\IUCrOrcidlink{0000-0002-3598-6880}}%
\author[a]{Erik Brambrink\IUCrOrcidlink{0009-0004-2404-9412}}%
\author[f]{Fabien Brieuc}%
\author[g]{Byoung Ick Cho\IUCrOrcidlink{0000-0003-1761-1150}}%
\author[a]{Sebastian G\"{o}de\IUCrOrcidlink{0009-0006-5266-9154}}%
\author[c]{Hauke H\"{o}ppner\IUCrOrcidlink{0009-0000-1929-5097}}%
\author[h]{Gerhard Jakob\IUCrOrcidlink{0000-0001-9466-0840}}%
\author[h]{Mathias Kl\"{a}ui\IUCrOrcidlink{0000-0002-4848-2569}}%
\author[a]{Zuzana Kon\^{o}pkov\'{a}}%
\author[g]{Changhoo Lee\IUCrOrcidlink{0009-0009-4862-7906}}%
\author[g]{Gyusang Lee}%
\author[a]{Mikako Makita\IUCrOrcidlink{0000-0003-1513-9198}}%
\author[a]{Mikhail Mishchenko\IUCrOrcidlink{0000-0002-2362-9392}}%
\author[j]{Mianzhen Mo\IUCrOrcidlink{0000-0002-2962-0815}}%
\author[d]{Pascal D. Ndione\IUCrOrcidlink{0000-0001-5129-0464}}%
\author[c]{Franziska Paschke-Bruehl\IUCrOrcidlink{0009-0001-0047-1051}}%
\author[k]{Michael Paulus\IUCrOrcidlink{0000-0002-3409-6798}}%
\author[c]{Alexander Pelka\IUCrOrcidlink{0009-0001-3308-5376}}%
\author[a]{Thomas R. Preston\IUCrOrcidlink{0000-0003-1228-2263}}%
\author[l]{Christian R\"{o}del}%
\author[c]{Michal ~\v{S}m\'{\i}d\IUCrOrcidlink{0000-0002-7162-7500}}%
\author[j]{Ling Wang}%
\author[a]{Lennart Wollenweber\IUCrOrcidlink{0000-0002-6934-1108}}%
\author[d]{Baerbel Rethfeld\IUCrOrcidlink{0009-0008-9921-4127}}%
\author[a]{Jan-Patrick Schwinkendorf\IUCrOrcidlink{0009-0002-8703-7641}}%
\author[b]{Christian Gutt\IUCrOrcidlink{0000-0002-0051-8542}}%
\author[a]{Motoaki Nakatsutsumi\IUCrCemaillink{motoaki.nakatsutsumi@xfel.eu}\IUCrOrcidlink{0000-0003-0868-4745}}%

\affil[a]{European X-ray Free-Electron Laser Facility, Holzkoppel 4, 22869 Schenefeld, Germany}

\affil[b]{Department Physik, Universit\"{a}t Siegen, 57072, Siegen, Germany}
\affil[c]{Helmholtz-Zentrum Dresden-Rossendorf, 01328, Dresden, Germany}

\affil[d]{Department of Physics and Research Center OPTIMAS, RPTU University Kaiserslautern–Landau, 67663 Kaiserslautern, Germany}

\affil[e]{Technical University Dresden, 01069 Dresden, Germany}

\affil[f]{CEA, DAM, Bruy\`eres-le-Ch\^atel, 91297 Arpajon, France}

\affil[g]{Department of Physics and Photon Science, Gwangju Institute of Science and Technology (GIST), Gwangju 61005, Republic of Korea}

\affil[h]{Institute of Physics, Johannes Gutenberg-Universit\"{a}t Mainz, 55128 Mainz, Germany}

\affil[j]{SLAC National Accelerator Laboratory, Menlo Park, CA 94025, USA}

\affil[k]{Fakult\"{a}t Physik/DELTA, TU Dortmund, 44221 Dortmund, Germany}

\affil[l]{Hochschule Schmalkalden, 98574 Schmalkalden, Germany}

\affil[$\dagger$]{present address: Department Physik, Universit\"{a}t Siegen, 57072, Siegen, Germany}

\begin{document} 
\maketitle

\begin{abstract}
We demonstrate a grazing-incidence x-ray platform that simultaneously records time-resolved grazing-incidence small-angle x-ray scattering (GISAXS) and grazing-incidence x-ray diffraction (GID) from a femtosecond laser–irradiated gold film above the melting threshold, with picosecond resolution 
\revision{using} an x-ray free-electron laser (XFEL). By tuning the x-ray incidence angle, the probe depth is set to tens of nanometers, enabling depth-selective sensitivity to near-surface dynamics. GISAXS resolves ultrafast changes in surface nanomorphology (correlation length, roughness), while GID quantifies subsurface lattice compression, grain orientation, melting, and recrystallization. The approach overcomes photon-flux limitations of synchrotron grazing-incidence geometries and provides stringent, time-resolved benchmarks for complex theoretical models of ultrafast laser–matter interaction and warm dense matter. Looking ahead, the same depth-selective methodology is well suited to inertial confinement fusion (ICF): it can visualize buried-interface perturbations and interfacial thermal resistance on micron to sub-micron scales that affect instability seeding and burn propagation.
\end{abstract}


\section{\label{sec:Introduction}Introduction}
\vspace{5pt}

Grazing-incidence (GI) x-ray techniques offer a unique capability to probe structural dynamics at surfaces and buried interfaces with high spatial resolution~\cite{Tolan1999, Roth16, Hexemer2015AdvancedAnalysis}. By varying the incidence angle $\alpha_i$, one can control the effective probe depth, thereby achieving depth-selective sensitivity that is unavailable in conventional transmission-based geometries. In the grazing-incidence small-angle x-ray scattering (GISAXS) and grazing-incidence x-ray diffraction (GID, or grazing-incidence wide-angle x-ray scattering: GIWAXS) combination, nanomorphology (roughness, ripples, correlation lengths), and subsurface lattice response (strain, disordering, recrystallization) can be captured simultaneously~\cite{Perlich2010, Richard2006_GID_GISAXS, Hexemer2015AdvancedAnalysis, Martin17_GISAXS_GIWAXS_Au}. GI x-ray methods have, to date, been most widely applied at synchrotron facilities. In this environment, obtaining sufficient scattering signal typically requires temporal integration on the microsecond to millisecond scale, which precludes direct access to truly ultrafast dynamics. Consequently, prior GI studies have predominantly targeted comparatively slow processes (e.g., growth, annealing, diffusion, and electrochemical cycling). Recently, we demonstrated that extending GISAXS to x-ray free-electron lasers (XFELs) overcomes the photon-flux bottleneck and enables visualization of embedded nanometric multilayer deformation and mixing, as well as the evolution of surface roughness and lateral correlation at sub-picosecond (ps) and nanometer (nm) scales~\cite{Randolph22, Randolph24}. This advance converts what previously required long integrations into a single-shot femtosecond x-ray measurement. 

This surface-sensitive, depth-selective approach addresses pressing needs across high-energy-density (HED) sciences. These include: (1) ultrafast laser-matter interaction at solid surfaces (e.g., melting kinetics, warm dense matter formation, laser ablation, and micromachining) and (2) inertial confinement fusion (ICF), where density/temperature modulations at buried interfaces seed hydrodynamic instabilities. In both cases, the key physics unfold within the first few tens of nm beneath a surface or interface and evolves on ps to nanosecond (ns) scales. Thus, a technique that couples nm-scale depth sensitivity with ps temporal precision, while simultaneously resolving lateral order, can bridge a long-standing diagnostic gap between optical pump-probe approaches (limited depth and sub-micrometer sensitivity) and post-mortem microscopy (no dynamics).

For ultrafast processing of metals, femtosecond (fs) optical excitation deposits energy into conduction electrons within the optical depth, followed by ultrafast electron-electron equilibration and electron-phonon coupling that drive heat and pressure into the lattice. Depending on the material, hot electrons can ballistically transport the laser energy into larger depths \cite{Hohlfeld1997, Byskov-Nielsen2011}, which is often described in terms of an \textit{effective} 
energy penetration depth~\cite{Hohlfeld2000,Ivanov2009,Rethfeld2017}. This nonequilibrium pathway launches coherent acoustic wave and, above threshold, induces localized melting confined to the top tens of nm on ps-ns scales~\cite{Rethfeld2017, Zhigilei09}; at higher fluence, stress confinement and rapid decompression can produce spallation or phase explosion~\cite{Zhigilei09,Shugaev2016,Sun25_thinAu_XFEL}. 
The interplay of electron heat capacity, heat transport and electron-phonon coupling determines the heating rate and the spatial temperature profile. Further, hydrodynamics and feedback between absorption and topography underlie the formation of laser-induced periodic surface structures (LIPSS)~\cite{Rudenko20, Bonse20}, which allows optimizing material properties including wettability, optical absorption, and tribological performance~\cite{Vorobyev2013, Lutey2018, Bonse_Tribology18}. Despite recent theoretical 
advances~\cite{Rudenko20, Terekhin2020, Nakhoul21selfOrganization, Zhang23Nanostructuring}, a direct experimental verification of the underlying ultrafast dynamics has remained elusive due to the lack of techniques that combine nm spatial resolution with ps temporal precision. Most studies rely on post-mortem SEM/AFM analysis or optical pump–probe methods~\cite{Hoehm_fsLIPSS_13, Garcia-Lechuga_psLIPSS_16, Terekhin2022}. Although recent transmission-based small/wide angle x-ray scattering (SAXS/WAXS) experiments at XFELs have resolved melt-front evolution, ripple formation, and spallation or phase-explosion signatures~\cite{Bonse24_SAXS_LIPSS, Sun25_thinAu_XFEL}, such measurements integrate over thickness and struggle to disentangle depth-dependent responses in bulk-like targets. A grazing-incidence approach, by contrast, can isolate the evolving near-surface layer where energy is first injected and where ripple/roughness development couples strongly to subsurface melting. 

In ICF, recent implosion experiments have reached the burning-plasma/ignition regime, achieving megajoule-scale yields with capsule gains exceeding unity\revision{~\cite{AbuShawareb_PRL24_ICF, Kritcher24_PRE_ICF}}. In this regime, small density/temperature discontinuities at buried interfaces, and possible interfacial thermal resistance, can influence burn propagation and symmetry~\cite{Craxton15_directICFreview, Allen25_ICF}. However, direct \textit{in situ} evidence of such buried modulations at relevant $\si{\micro\meter}$ resolution has been scarce. Hydrodynamic theory and experiments emphasize that ablative stabilization suppresses only sufficiently short wavelengths, leaving a band of micron-to-tens-of-micron modes that can seed Rayleigh–Taylor growth~\cite{Craxton15_directICFreview, HAMMEL10_RTI}.
These considerations motivate a surface \revision{ and interface}-sensitive, depth-selective x-ray probe \revision{
that can quantify the early-time interfacial seed spectrum and its evolution. Grazing-incidence scattering directly accesses in-plane length scales \( L = 2\pi/Q_{x,y} \), spanning approximately nanometers to several micrometers (depending on the accessible Q-range and detector geometry), and can follow their evolution from picoseconds into the nanosecond regime (by changing the pump-probe delay). For accessing spatial length scales beyond several micrometers, grazing-incidence x-ray imaging would be a more appropriate approach ~\cite{Fenter06_GIXI}; however, such capabilities have not yet been demonstrated at XFELs.}

In this work we therefore deployed time-resolved GISAXS+GID at an XFEL as a general platform for HED interfacial dynamics. 
We chose here ultrafast laser excitation of metal films as a demonstrative case: it exercises the same measurement capabilities (depth selectivity, lateral sensitivity, and ps timing) that are required for ICF-relevant interfaces, yet in a more compact setting. The combined GISAXS/GID observables directly correlate lateral nanostructure with subsurface lattice evolution within the x-ray penetration depth, providing stringent benchmarks for multiscale models of laser ablation and, by extension, for the interfacial transport and stability physics that govern ICF performance.

\begin{figure*}[hbt!]
\centering
\includegraphics[width=1\linewidth]{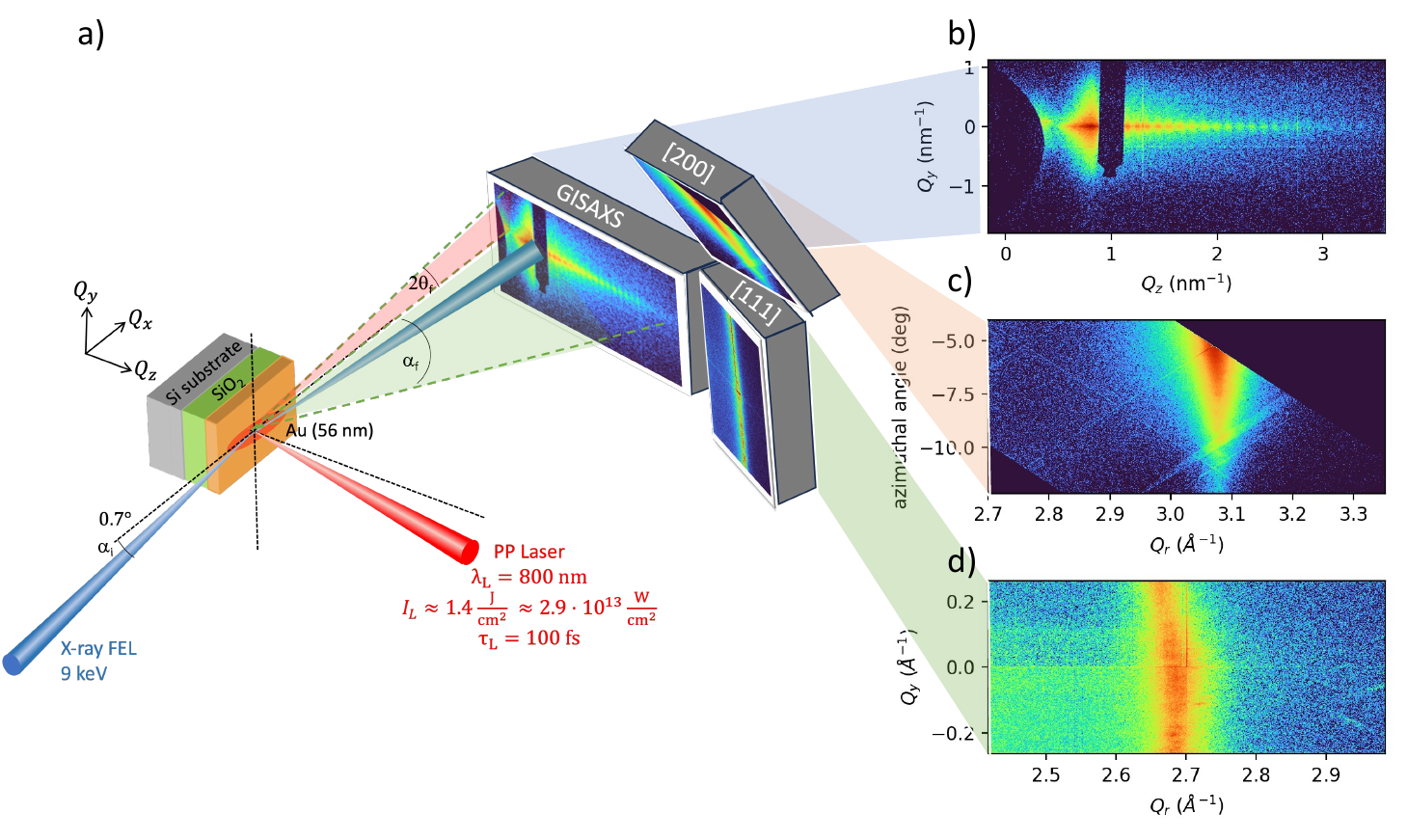}
     \caption{(a) Schematic representation of the experimental setup. Horizontally polarized \(\mathrm{h\nu} = 9\,\mathrm{keV}\) x-ray pulses are directed onto the thin film sample at a grazing incidence angle \(\alpha_i = 0.7 \degree \), slightly above the critical angle for total external reflection \(\alpha_c = 0.495 \degree \). This limits the x-ray penetration depth to be \( \SI{30} {\nano\meter} \) for gold. The sample surface is oriented perpendicular to the ground. A 2D GISAXS detector captures scattering in the small-angle reflection geometry, defined by the exit angle \(\alpha_f\) (\(x-z\), in-plane scattering, perpendicular to the sample surface) and \(2\theta_f\) (\(x-y\) plane, out-of-plane scattering, parallel to the sample surface). Additionally, two 2D detectors are placed in the wide-angle scattering region to serve as grazing-incidence diffraction (GID) detectors. 
     (b) Typical GISAXS pattern from an x-ray only shot (cold sample). The fringes along $Q_z$ are used to determine the sample thickness. The GID patterns for the (c) (200) and (d) (111) peaks of Au thin film observed $\SI{30}{\pico\second}$ after laser irradiation.}      
\label{fig:ExpSetup} 
\end{figure*}

\section{\label{sec:Exp_setup}Experimental setup}
\vspace{5pt}

\subsection{X-ray, laser and sample}
The experiment was conducted at the HED/HiBEF instrument at the European XFEL facility~\cite{zastrau21HED_JSR}. The experimental setup is schematically shown in Fig.~\ref{fig:ExpSetup}. A $\SI{56}{\nano\meter}$ thick gold (Au) film was deposited on a \( \SI{100}{\nano\meter} \) thermal oxide (\( \mathrm{SiO_2} \)) layer grown on a Si $\SI{700} {\micro\meter}$ thick substrate. 
The samples were irradiated by a pump-probe (PP) laser~\cite{Palmer19_PPL} at close-to-normal incidence, with a central wavelength of \(\SI{800}{\nano\meter} \), a maximum pulse energy of $\SI{700}{\micro J}$, and a pulse duration of $\approx\,\SI{100}{\femto\second}$ full-width at half maximum (FWHM). The sample was then probed with an x-ray pulse of \( \approx \, \SI{20} {\femto\second} \) duration at 9.0~keV photon energy, in a grazing-incidence geometry with an incidence angle of \( \alpha_i = \SI{0.7}{\degree} \), corresponding to a penetration depth of \( \SI{30}{\nano\meter} \). Thus, the measurement is strongly surface sensitive and comparable to the optical skin depth of gold,  $\SI{13.5}{\nano\meter}$ at $\lambda = \SI{800}{\nano\meter}$. The x-ray beam was focused using a beryllium compound refractive lens (CRL) assembly positioned about 30~cm upstream of the sample. A point projection of a Siemens star at the focal plane indicated an x-ray spot size of approximately $\SI{1}{\micro\meter}$, which is limited by the lens chromaticity and the SASE (Self-Amplified Spontaneous Emission) bandwidth ($\approx$ 30 eV). This resulted in an approximately $\SI{80}{\micro\meter}$ footprint on the sample in the horizontal direction. As one edge of the x-ray beam arrived on sample earlier than the other edge, the footprint determines the temporal integration of $\SI{270}{\femto\second}$. In order to mitigate the laser intensity variation along the probing area, the laser is defocused with respect to the sample position, to achieve a laser beam size of approximately $200 \times \SI{90}{\micro\meter\squared}$ (FWHM), elongated in the horizontal (= x-ray footprint) direction. Assuming 30 \% encircled energy, the maximum laser fluence on the sample is $ \SI{1.4} {\joule\per\centi\meter\squared}$, corresponding to an intensity of $ \SI{2.9e13}{\watt\per\centi\meter\squared}$. Achieving a good spatial overlap between the laser and x-ray beam is one of the most challenging aspects of this type of experiment, as a misalignment of \(\le \SI{2} {\micro\meter}\) along the sample surface normal ($z$-direction) would be enough to lose the overlap. We performed post-mortem analyses using an optical microscope and selected data where damage from both laser and x-rays clearly overlapped on the sample. This analysis also revealed that pulse-to-pulse pointing fluctuations of both x-rays and laser were negligible, confirming that the loss of spatial overlap is solely caused by the sample positioning accuracy. For the x-rays, thanks to the final focusing optics placed near the sample, lateral pointing fluctuations are converted into transmission loss (due to a mismatch between the optical axis of the lens and the x-ray axis), while leaving the focus position unaffected. This also implies that the x-ray intensity on the sample fluctuated by pointing variations upstream of the lens. Since we did not have a monitor to measure shot-to-shot CRL transmission, nor saved the data on x-ray pointing before the lens, the x-ray data shown in this paper are \textit{not} properly normalized in intensity -- they are only normalized to the x-ray intensity gas monitor (XGM)~\cite{XGM_Maltezopoulos2019} placed upstream of the lens. 
The synchronization between the x-ray and the laser pulse at the sample location was achieved using a cerium-doped yttrium aluminum garnet (Ce:YAG) scintillation crystal, with a temporal accuracy of about $\SI{100}{\femto\second}$. The root-mean-square (RMS) pulse-to-pulse temporal jitter was approximately $\SI{30}{\femto\second}$ which is much smaller than the temporal dynamics we discuss in this paper. The delay between the x-ray and the laser is achieved by moving a delay line of the optical laser. \revision{The sophisticated timing infrastructure at FEL facilities enables pump–probe delay scans spanning a few fs to many ns (and beyond) by combining optical delay lines with RF (radio-frequency) phase control and programmable trigger timing.~\cite{Schulz14thesis}}

\begin{figure*}[hbt!]
    \centering
    \includegraphics[width=1\linewidth]{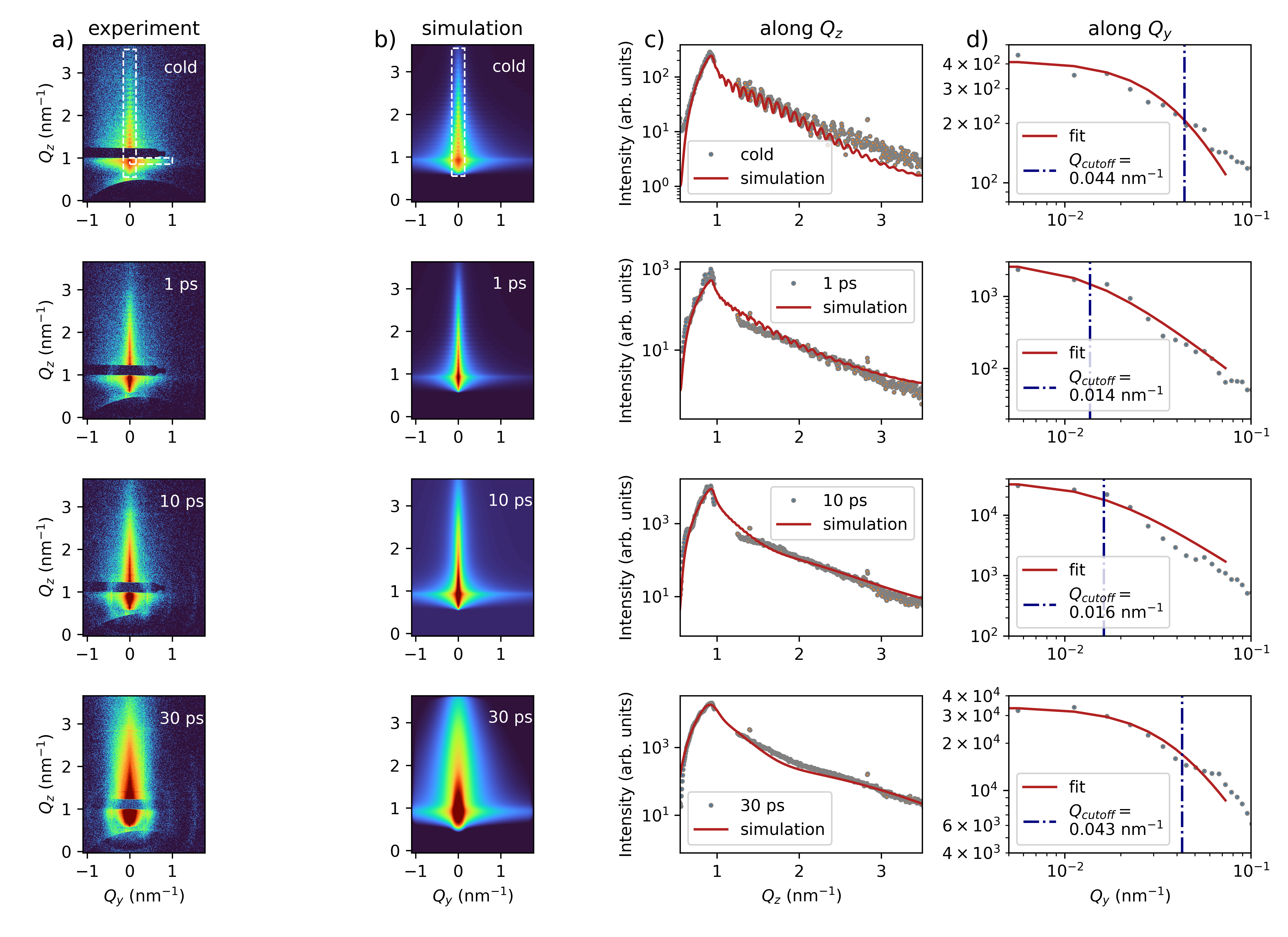}
    \caption{\label{fig:GISAXS_all} 
    Time-resolved GISAXS analysis of the Au sample at different time delays. Panel (a) shows the experimental GISAXS data obtained at the cold reference (unexcited state), as well as at 1, 10, and 30 ps after laser irradiation. \revision{The white dashed boxes indicate the regions taken for lineouts.} Panel (b) presents the corresponding simulations of the experimental conditions, illustrating the modeled evolution of the system over time. Panel (c) displays the scattering profiles for the in-plane ($Q_z$) direction comparing both experimental and simulated results. Panel (d) shows the scattering profiles for the out-of-plane ($Q_y$) direction. The red line represents the model used to determine the cutoff value, while the blue vertical line indicates the position of this cutoff. 
    }
\end{figure*}

\begin{table*}[hbt!]
    \centering
    \caption{\label{table: BornAgain} 
    Parameters of vertical correlation length ($\xi_\perp$), thickness ($d$), roughness ($R$) for Au and Si \revision{(retrieved from BornAgain simulation)} and refined $Q_\text{cutoff}$ at various delays.}      
    \begin{tabular}{|c|c|c|c|c||c|}
    \hline
    Delay (ps) & $\xi_\perp$ (nm) & $d_{\mathrm{Au}}$ (nm) & $R_{\mathrm{Au}}$ (nm) &  $R_{\mathrm{Si}}$ (nm) & $Q_{\mathrm{cutoff}}$ (nm$^{-1}$) \\
    \hline
    0  & 26 & 56.4 & 0.9  & 0.4  &  0.044 \\
    1  & 21 & 56.4 & 1.0  & 0.4  &  0.014\\
    10 & 3  & 56.4 & 1.4  & 0.4  &  0.016\\
    30 & 1  & 49.0   & 1.3  & 0.5  &  0.043\\
    \hline
    \end{tabular}
     \label{tab:table1}
\end{table*}

\subsection {Grazing-incidence small-angle x-ray scattering (GISAXS)}
GISAXS measurements were conducted to enhance surface sensitivity while simultaneously probing the in-plane ($Q_x$,  $Q_z$) and out-of-plane ($Q_y$) nanostructural features of the sample. For clarity, we note that the terms 'in-plane' and 'out-of-plane' in this work refer to the scattering geometry, i.e. the plane defined by the incident and scattered x-ray wavevectors, and not to the crystallographic or surface plane of the sample. The incident x-ray beam was aligned at a small angle  ($ \alpha_i $) relative to the sample surface, close to the critical angle of total external reflection ($\alpha_c$), to minimize bulk scattering and optimize surface-sensitivity.
The scattered intensity consists of two main components: specular reflection and diffuse scattering. The specular reflection occurs when the incident and exit angles are equal ($\alpha_i = \alpha_f$ and $2\theta_f=0$) and is usually several orders of magnitude stronger than the diffuse scattering in the chosen scattering geometry. The diffuse scattering arises from surface or interface roughness, leading to scattered intensity distributions around the specular peak. This component is sensitive to lateral correlations and periodicities in the surface or interface morphology. We blocked the intense specular peak to prevent detector saturation and focused on analyzing the diffuse scattering. The signal was recorded using a charge-integrating JUNGFRAU pixel detector~\cite{mozzanica18_JF}.

The scattered intensity is characterized by the momentum transfer components:
\begin{align}
Q_x &= \frac{2\pi}{\lambda}\left(\cos(\alpha_f)\cos(2\theta_f)-\cos(\alpha_i)\right) \\
Q_y &= \frac{2\pi}{\lambda}\left(\cos(\alpha_f)\sin(2\theta_f)\right) \\
Q_z &= \frac{2\pi}{\lambda}\left(\sin(\alpha_f)+\sin(\alpha_i)\right). 
\end{align}
Here, $\alpha_i$ denotes the fixed incident angle, $\alpha_f$ describes the exit angle within the scattering plane (\textit{in-plane}, $x-z$) and $\theta_f$ is the exit angle perpendicular to the scattering plane (\textit{out-of-plane}, $x-y$) as indicated in Fig.~\ref{fig:ExpSetup}.
Although surface and interface roughness contributes to both the in-plane ($Q_z$) and the out-of-plane ($Q_y$) scattering components, $Q_z$ primarily encodes vertical structural features such as layer thickness, whereas $Q_y$ reflects lateral correlations and structural periodicities within the sample plane.

As seen in Fig.~\ref{fig:GISAXS_all}, a peak in \(Q_z\) appears at exit angles below the incident angle ($Q_z < Q_{\text{specular}}$). This intensity enhancement occurs when the exit angle of the scattered x-rays matches the material's critical angle for total external reflection, creating a resonance of the evanescent x-ray wave. This so-called Yoneda peak, caused by interference in the topmost surface layers, is highly sensitive to the refractive index contrast between the sample and its surroundings, providing insights into surface roughness and  near-surface structure. The Yoneda peak appears at \(\alpha_f = \alpha_c = \sqrt{2\delta}\), where \( \alpha_c \) is the critical angle for total external reflection, \(1 - \delta\) is the real part of the refractive index and \(\delta \propto \rho_e \), where \(\rho_e\) is the electron density. For gold at \(h\nu = 9 \) keV, the critical angle is at $\alpha_c = 0.495 \degree$, which corresponds to \(Q_z = 0.95 ~\mathrm{nm}^{-1} \).

\subsection {Grazing-incidence diffraction (GID) from a textured sample}
 In order to correlate the surface morphology inferred by GISAXS and the underlying subsurface crystallographic changes, two dedicated diffraction detectors (ePix hybrid-pixel detector [27] and JUNGFRAU detector [26]) were placed at the diffraction peaks (111) and (200) for gold. 
Our magnetron sputtered gold sample, deposited \revision{on 100 nm thermally oxidized SiO$_2$ grown} on (100)--oriented silicon wafer, exhibits a preferred grain orientation along a single axis (fiber texture). 
Specifically, the reciprocal lattice vector $G_{111\_\mathrm{cold}}$ is preferentially aligned with the $z$-axis (normal to the sample surface), with a rocking curve width of approximately $5\degree$. 
In typical transmission x-ray diffraction geometry, this texture causes a strong azimuthal dependence in the diffraction pattern \cite{McGonegle15}. Under grazing incidence geometry, diffraction from certain lattice planes is suppressed in textured samples. For example, at $h\nu = 9 \,\mathrm{keV}$, the Bragg angle from the (111) plane is \(2\theta_{111} = 34 \degree\) (corresponding to \(2.67\text{\AA}^{-1}\)), assuming a lattice constant of $a = 4.08 \, \text{\AA}$. Therefore, the Bragg condition for the (111) plane (where the incident and outgoing angles are equal) cannot be satisfied, resulting in the absence of this peak. Reflections from other planes, however, may still appear at specific positions on the Debye-Scherrer ring. For instance, the (200) plane, which forms an angle of $\gamma = 54.7^\circ$ with the (111) plane, is visible in certain directions (see Methods). Its diffraction intensity is expected to decrease as the sample texture is lost due to, \textit{e.g.,} laser excitation, offering insight into the lattice response along this orientation.
Conversely, at the (111) peak position located at \(2\theta_{111}\) in the \(x-z\) plane (perpendicular to the sample surface), diffraction signal is expected to appear only after the sample becomes disordered. The broad peak of the liquid structure factor \(S(Q)\) of gold would appear slightly below the $Q_{111}$ peak due to volume expansion from solid to liquid.

\section{\label{sec:Results} Results and discussions}
\vspace{5pt}
\begin{figure*}[hbt!]
    \centering
    \includegraphics[width=1\linewidth]{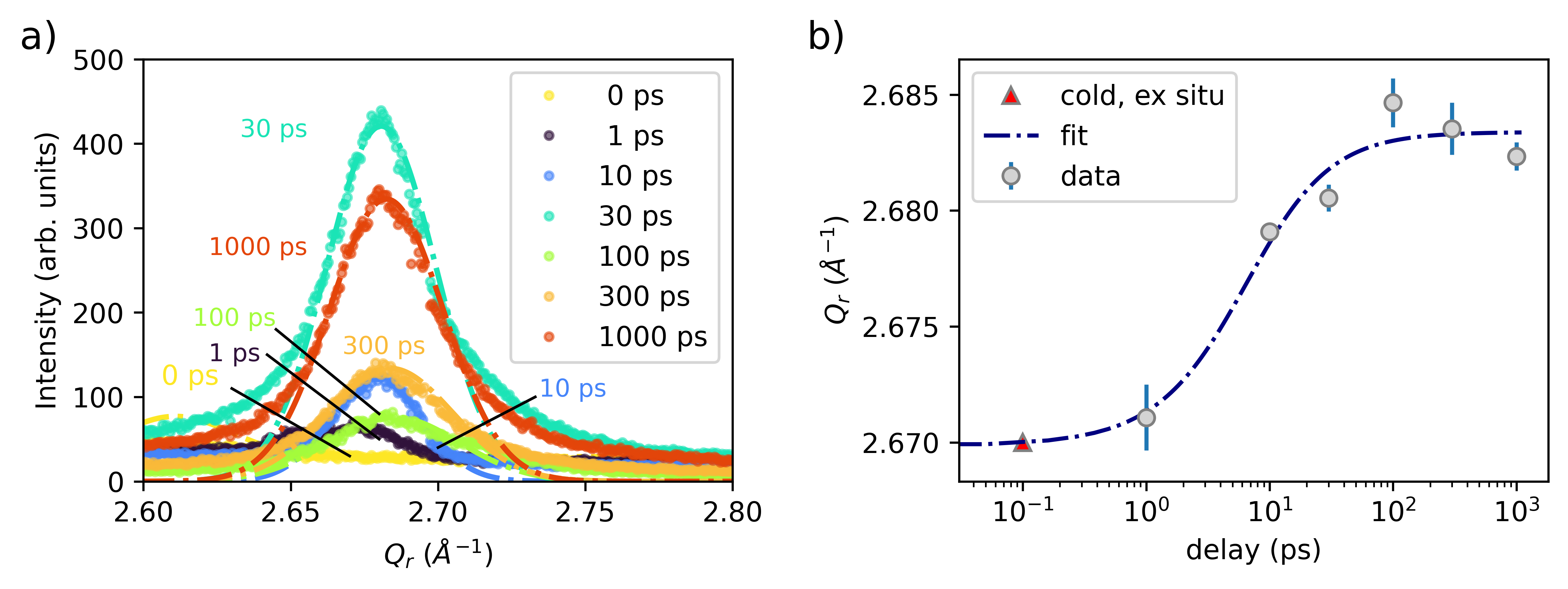}
    \caption{\label{fig:Au111} 
    (a) Temporal evolution of the Au (111) diffraction peak following laser excitation at various time delays. The absence of the Au (111) peak in the cold sample is attributed to the initial sample texture along the surface. Upon laser excitation, the Au (111) peak emerged, likely due to grain redistribution associated with surface melting and subsequent recrystallization. The dashed lines show a gaussian fit. Note that the peak intensity is \textit{not} correctly normalized, due to fluctuations in CRL transmission caused by beam pointing instability. Here, $Q_r$ is defined as $\sqrt{Q_y^2+Q_z^2}$. (b) Evolution of the Au (111) peak positions over time reveals compression along this plane. Note that the (111) peak appears at \(\sim 2.67 \text{\AA}^{-1}\) (lattice constant  \(4.08 \text{\AA}\)) in the cold sample, as characterized \textit{ex situ} using an x-ray diffractometer.
    }
\end{figure*}

\begin{figure*}[hbt!]
    \centering
    \includegraphics[width=1\linewidth]{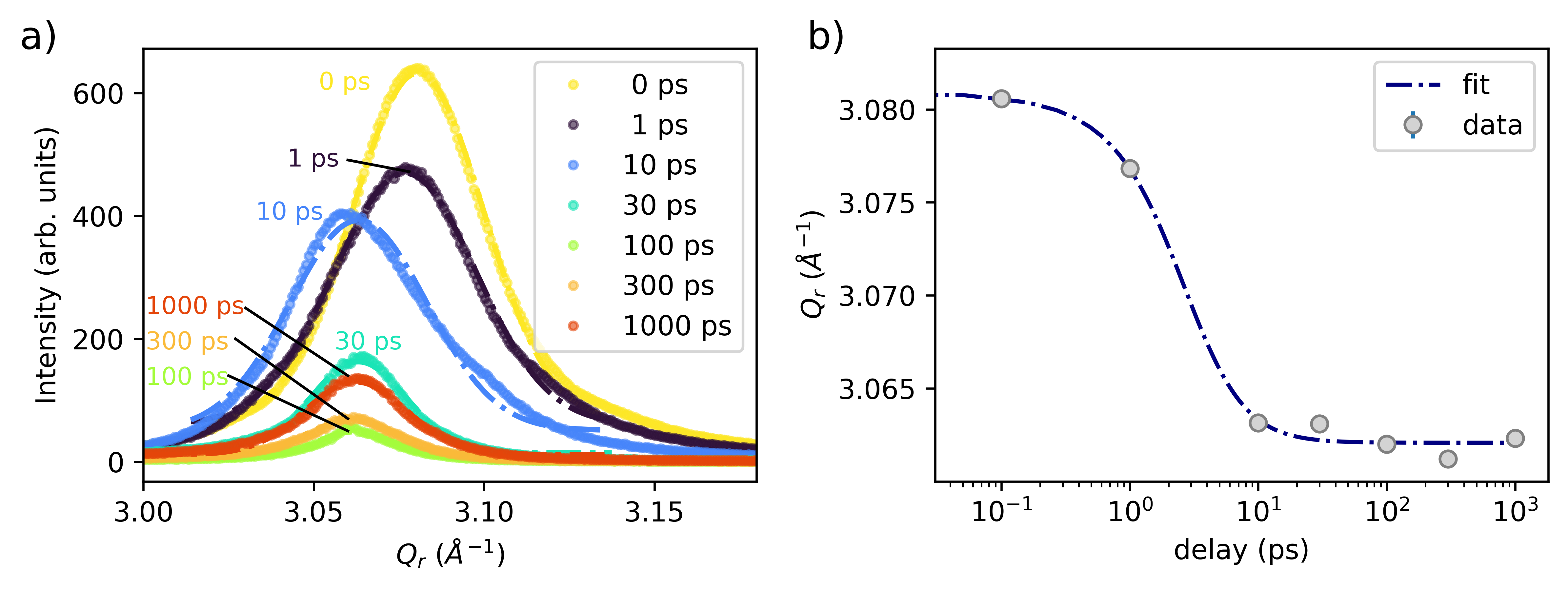}
    \caption{\label{fig:Au200} 
    (a) Temporal evolution of the Au (200) diffraction peak following laser excitation at various time delays. The detector position for the (200) reflection corresponds to the (111) fiber texture, which exhibits a peak width of $\sim 5\degree$ as seen in Fig.~\ref{fig:ExpSetup} c) . Upon melting and recrystallization, this intensity becomes azimuthally dispersed and weakens, so the (200) reflection essentially probes the unmelted, cold region. The dashed lines show a gaussian fit. (b) Evolution of the Au (200) peak positions with time which reveals expansion along this plane. 
    }
\end{figure*}

Fig. \ref{fig:ExpSetup} b) shows a typical GISAXS pattern from a cold Au sample (prior to laser irradiation) providing insights into the surface and sub-surface structural properties of materials. The oscillations along $Q_z$ arise from interference between the free surface and the \( \mathrm{Au} / \mathrm{SiO_2} \) interface; their period yields a layer thickness of $\SI{56}{\nano\meter}$. 
Upon laser irradiation, we observe significant changes in the GISAXS patterns. Fig. \ref{fig:GISAXS_all} a) summarizes the data for the cold sample (top) and $1$, $10$ and $\SI{30}{\pico\second}$ after laser irradiation. The small oscillations along $Q_z$ gradually fade and vanish by $\approx\SI{10}{\pico\second}$, indicating the loss of cross-correlations in surface/interface morphology between the Au surface and the \( \mathrm{Au} / \mathrm{SiO_2} \) interface. Such ultrafast changes in surface morphology are indicative of melting. These features are reproduced using the open-source software BornAgain, based on the distorted-wave Born approximation \cite{Pospelov:ge5067} (Fig. \ref{fig:GISAXS_all} b)). 
We refined the BornAgain model to obtain relevant parameters, including layer thickness, roughness, and correlation lengths. The refinement yields an initial surface root-mean-square (RMS) roughness of $\SI{0.9}{\nano\meter}$, increasing to $\SI{1.4}{\nano\meter}$ upon laser excitation. Simultaneously, the vertical correlation length $\xi_\perp$ decreases from $26$ to $\SI{1}{\nano\meter}$. This parameter describes the average distance over which structural correlations persist along the sample depth direction (see Methods).
Fig. \ref{fig:GISAXS_all} c) compares experimental and simulated scattering lineouts along $Q_z$ (and $Q_x$) directions, which provides complementary information on vertical and lateral structural correlations. The retrieved parameters are summarized in Table \ref{table: BornAgain}. \revision{Meanwhile}, Fig. \ref{fig:GISAXS_all} d) displays the experimental scattering profiles along the $Q_y$ direction. Owing to the presence of multiple decay regimes along $Q_y$, the BornAgain simulations cannot fully reproduce the experimental data. To model the initial decay, we use an intensity dependence of the form $\propto \frac{1}{Q_y^2 + Q_\text{cutoff}^2}$, where $Q_\text{cutoff}$ denotes the shoulder position \cite{PhysRevB.38.2297} and is related to the lateral correlation length $\xi_\parallel$ \cite{Tolan1999} (see Methods); the refined $Q_\text{cutoff}$ is indicated by the blue vertical line.
After $\SI{1}{\pico\second}$ the cutoff position drops sharply from $Q_y = 0.044$ to $0.014$ nm$^{-1}$ (real space scales $2\pi/Q_y$ of $\approx 140$ to $\approx \SI{450}{\nano\meter}$). 
This abrupt decrease is consistent with the onset of surface melting, with $Q_\text{cutoff}$ moving beyond the accessible resolution range. In this regime, the scattering is likely governed by the transverse coherence length of the x-ray beam in combination with the angular resolution limit rather than intrinsic sample correlations~\cite{Tolan1999}. From $10$ to $\SI{30}{\pico\second}$, $Q_\text{cutoff}$ gradually returns towards its initial value, indicating the onset of resolidification. \revision{We note that the scattering profiles along $Q_y$ are extracted at the Yoneda peak, providing enhanced sensitivity to near-surface structural correlations. The rapid decrease of $Q_\text{cutoff}$ within the first picosecond therefore reflects a prompt loss of lateral correlations in the surface-near region, consistent with the onset of surface-localized melting. The simultaneous reduction of the $Q_z$ oscillations already after $\SI{1}{\pico\second}$ indicates that vertical correlations are reduced at similar times. Their complete disappearance by $\SI{10}{\pico\second}$ indicates that a sufficiently large fraction of the film thickness is molten or structurally disordered, leading to a full decoupling of the surface morphology from the buried interface.}
While GISAXS resolves nanometer-scale structural dynamics, it cannot capture subsurface atomic-scale changes. To probe the underlying atomic-scale processes, we now turn to the GID signal.

In Fig.~\ref{fig:Au111} a), lineouts of the Au (111) diffraction peak are shown for various time delays. No peak is observed in the unpumped (cold) sample, as expected, because at grazing incidence the reciprocal lattice vector is not aligned with the [111] direction. Upon laser excitation, the (111) peak emerges, and its intensity increases within 30~ps, indicating grain rotation. Such ultrafast grain redistribution is typically associated with melting~\cite{Li17}. Complete melting would eliminate the (111) peak and produce a broader and weaker liquid diffuse signal; in our case only a subtle increase in diffuse scattering around the (111) peak is observed, consistent with partial melting common to transient melt states~\cite{Robinson23, Descamps24}. The emergence of a well-defined (111) peak—initially absent in the cold state— therefore indicates recrystallization following melting, with randomly oriented grains producing an azimuthally homogeneous (111) reflection.
\revision{After $\SI{30}{\pico\second}$, the (111) reflection peak remains clearly observable at all later delays, despite variations in its measured intensity which we attribute to variations in the incident x-ray flux arising from pointing fluctuations upstream of the final focusing CRL optic.} These observations indicate surface-localized transient melting during the first few tens of ps. Once the surface temperature falls below the melting threshold, recrystallization begins at about 30 ps (for our fluence), consistent with Li et al. (Cu)\cite{Li17} and Zhigilei et al. (Ni)\cite{Zhigilei09} for absorbed fluences of a few tens of $\si{\milli\joule\per\centi\meter\squared}$. \revision{Here, Zhigilei et al. used 1 ps pulses, in contrast to the 100 fs pulses used in our experiment; although the pulse duration can affect transient absorption, the tens-of-ps resolidification timescale is expected to be set mainly by the post-equilibration temperature profile (i.e., the absorbed energy density). This is consistent with the 100 fs results of Li et al. (2017), which reported similar skin-depth-confined melting at the top few tens of nanometers and resolidification within a few tens of picoseconds near the fluence threshold for surface melting.} Fig.~\ref{fig:Au111} b) summarizes the temporal evolution of the Au(111) peak position obtained via Gaussian fitting. A rapid shift to higher $Q_z$ is observed within 1–10~ps, stabilizing after $\approx\SI{30}{\pico\second}$. The shift $(Q_{30ps}-Q_{0ps})/Q_{0ps} \sim 0.5 \%$ describes the strain along the reciprocal lattice vector $G_{111}$. In grazing incidence, $G_{111}$ makes an angle $\approx \theta_{111}$ to the sample normal, with $2\theta_{111} = \SI{34}{\degree}$ the Bragg angle. The projected strain is \(\epsilon_{\mathrm{proj}}(\theta_{111}) = \epsilon_{\perp} \cos^{2}\theta_{111} + \epsilon_{\parallel} \sin^{2}\theta_{111} = 0.92\,\epsilon_{\perp} + 0.085\,\epsilon_{\parallel}\), indicating predominant sensitivity to the strain along the surface normal.

The apparent \revision{longitudinal} compression is, at first glance, counterintuitive, since the melted \revision{near-}surface region \revision{would be expected to expand } 
at the free surface. 
\revision{Surface heating generates a compressive wave that propagates into the bulk, followed by an unloading (tensile) wave that promotes surface expansion. However, the (111) peak analyzed here arises primarily from randomly oriented grains after resolidification. 
The observed longitudinal compression relative to the initial lattice spacing is therefore more likely to reflect the post-melt stress/strain state than the transient acoustic response. A plausible explanation is elastic coupling via the Poisson effect: the observed longitudinal compression could result from lateral expansion. Because the gold film is laterally clamped by the thick substrate, residual lateral stress from sputter deposition is expected ~\cite{Faurie06} and can relax upon heating and melting. Consistent with this picture, we observe a pronounced lattice expansion with strong lateral sensitivity to the (200) reflection, as discussed below.
}

Fig.~\ref{fig:Au200} a) shows lineouts of the Au(200) diffraction peak at different delays; this peak is azimuthally confined \revision{in the cold state} due to the \{111\}-texture. Laser excitation induces lattice disordering and grain orientation, \revision{which reduces the (200)} diffraction intensity. Subsequent recrystallization does not restore the initial intensity because the initially azimuthally confined texture becomes azimuthally distributed.
\revision{
Fig.~\ref{fig:Au200} b) summarizes the evolution of the $Q_{200}$ peak position. The peak shift toward lower $Q$ with $(Q_{30ps}-Q_{0ps})/Q_{0ps} \approx -0.6\%$, indicating lattice expansion. Here, the reciprocal lattice vector $G_{200}$ makes an angle $\gamma = \SI{54.7}{\degree}$ with respect to the sample normal, such that \(\epsilon_{\mathrm{proj}\_200}(\gamma) = \epsilon_{\perp} \cos^{2}\gamma + \epsilon_{\parallel} \sin^{2}\gamma = 0.286\,\epsilon_{\perp} + 0.714\,\epsilon_{\parallel} \), making this measurement primarily sensitive to the lateral strain (parallel to the sample surface).
The observed lateral lattice expansion supports the hypothesis that the measured longitudinal compression could arise from Poisson coupling. Nevertheless, a quantitative estimates indicates that the Poisson coupling alone cannot fully account for the observed magnitude. 
Using a Poisson’s ratio of \( \nu = 0.42 \) for gold~\cite{Faurie06}, a longitudinal compression of 0.5 \% corresponds to a lateral tensile strain of only ~0.35 \%, which would yield \(\epsilon_{\mathrm{proj}\_200} = 0.4 \% \), smaller than the measured ~0.6 \% (see Methods).} 
\revision{
The remaining discrepancy may reflect that the (111) and (200) peaks are weighted toward different material fractions. The (111) peak predominantly reflects the recrystallized (previously melted) near-surface layer, whereas the (200) peak preferentially samples the remaining textured, less-disordered fraction. Near the melting threshold, the melt layer can remain confined to approximately the optical skin depth (here $\SI{\sim 13.5}{\nano\meter}$) [Li et al 2017], which is smaller than the x-ray penetration depth ($\SI{\sim 30}{\nano\meter}$). Therefore, the probed volume can include both resolidified near-surface material and relatively cooler/unmelted material. Here, heating-induced defects or voids in the less-disordered region could modify the effective density and lattice parameter and thereby contribute to the observed (200) peak shift.}
At this point, however, the discussion remains speculative within the limits of the current dataset. We emphasize that the experimental setup employed here enables simultaneous access to both depth-dependent melting dynamics and anisotropic strain evolution. 
\revision{
For instance, adding an additional detector at $\alpha_f \approx 0$ and $2\theta_f \approx \theta_{111}$ or $\approx\theta_{200}$, together with systematic variation of the incident angle to tune the penetration depth, would enable a more direct determination of the lateral lattice parameter and provide a more stringent test of the scenarios discussed above.
}

The absorption of \( \SI{800}{\nano\meter} \) light in gold and the subsequent temperature evolution are simulated by coupling a two-temperature model with the temperature-dependent Drude-Lorentz dielectric function~\cite{Ndione2024} (see Methods). For our incident laser fluence of \SI{1.4}{\joule\per\centi\meter\squared}, the model yields an absorbed fluence of approximately \SI{102}{\milli\joule\per\centi\meter\squared}.
The simulations indicate that the electron temperature in the gold film increases rapidly to greater than \(15 000 \, \mathrm{K} \) during the laser pulse. 
Energy is then transferred to the lattice on a time scale of roughly \SI{40}{\pico\second}.
Neglecting further dissipation processes, a final lattice temperature of about \( 6000 \, \mathrm{K} \) is reached within the \SI{56}{\nano\meter} film -- well above the equilibrium melting temperature of gold (\( T_m = 1338 \, \mathrm{K} \)). 
While this superheating appears to be rather high, the connection between equilibrium melting temperature and nucleation kinetics of a highly nonequilibrium state is much more involved \cite{Zhigilei09,Rethfeld2017}. 
Recent work reports about direct measurements of the lattice temperature of gold, which exceed its equilibrium melting temperature by far before complete melting \cite{White2025nature}.
On the other hand, there are uncertainties in the experimental fluence calibration; additional loss channels (e.g, heat flow due to the thick substrate); grain-boundary scattering that confines energy deposition near the surface area~\cite{Assefa20}; and any other effects not accounted for in the simulations.
Note that a measured incident fluence of \SI{4.2}{\joule\per\centi\meter\squared} for a film of twice the thickness compared to our work has been found to compare well with a molecular dynamic simulation applying an absorbed fluence of \SI{100}{\milli\joule\per\centi\meter\squared}~\cite{Sun25_thinAu_XFEL}. 
Moreover, inline with the discussion in Ref.~\cite{Sun25_thinAu_XFEL} regarding the uncertainty of the theoretical fluence conversion,
our simulations revealed a nonlinear increase of the absorbed fluence with the incident fluence, because the optical constants evolve during the pulse due to electronic excitation. 
Further experimental investigations will be required to identify the dominant mechanisms behind this discrepancy and to benchmark them against theoretical predictions.


\section{Conclusion and outlook} 
In this paper, we have demonstrated picosecond-resolved, simultaneous measurement of surface nanomorphology and subsurface atomic structure. By combining intense femtosecond XFEL pulses with GISAXS and GID, we quantitatively extract changes in surface nanostructure (e.g., correlation length and roughness) while concurrently resolving atomic-scale dynamics in the subsurface layer, including lattice disordering, melting, compression, expansion, and recrystallization.
Because the grazing-incidence configuration is experimentally demanding\revision{,} most notably the stringent requirement for x-ray/laser spatial overlap\revision{,} the analyzable dataset in this first campaign was limited. In addition, the absence of an on-sample x-ray intensity monitor precluded a fully quantitative analysis of diffraction-intensity dynamics. These limitations have now been addressed by implementing a surface-imaging camera for overlap verification and an on-sample x-ray monitor; data acquired with the improved platform will be reported in future work. Nevertheless, we emphasize that our setup enables simultaneous access to surface nanomorphology, depth-dependent lattice dynamics and anisotropic strain evolution, paving the way to elucidate laser-ablation physics and to benchmark theoretical models.
In forthcoming studies, varying the x-ray incidence angle — for example, to $\SI{1.7}{\degree}$ and $\SI{3.5}{\degree}$ for gold — will tune the probe depth to approximately $\SI{100}{\nano\meter}$ and $\SI{200}{\nano\meter}$, respectively. This depth-resolved capability will allow us to track e.g., thermal diffusion and melt-front propagation, that underlie laser-induced surface nanostructuring, including nanoscale periodic ripple formation~\cite{Rudenko20}. In addition, as dedicated HED platforms mature, the same grazing-incidence, time-resolved GISAXS/GID methodology can be applied to inertial confinement fusion (ICF) research. 

\section{METHODS} 
\subsection{Multilayer sample}
\noindent The multilayer (ML) sample was prepared by DC magnetron sputtering at the University of Mainz. 
The gold sample layers were deposited on a 100~nm thermal oxide \( \mathrm{SiO_2} \) layer grown on a \( \SI{700}{\micro\meter} \) thick silicon substrate. The wafer was then laser-cut into $25 \times 7 \,\mathrm{mm^2}$ individual pieces. The 100~nm thermal oxide layer provides a smoother surface, and its thickness was verified by x-ray reflectometry prior to coating.  

\subsection{Correlation functions}

To describe the lateral distribution of the surface morphology, the out-of-plane diffuse scattering along the $Q_y$ direction can be analyzed using the height--height correlation function:
\begin{equation}
C(R) = \langle h(0) h(R) \rangle = \sigma^2 \exp \left( - \left[ \frac{R}{\xi_\parallel} \right]^{2H} \right),
\end{equation}
where $R$ is the spatial separation, $\sigma$ is the root mean square (RMS) roughness, $\xi_\parallel$ is the lateral correlation length, and $H$ is the Hurst parameter~\cite{PhysRevB.38.2297}.

The correlation function along the sample depth $Q_z$ is given by:
\begin{align}
\langle h_j(0) h_k(R) \rangle = & \frac{1}{2} \left[ \frac{\sigma_k}{\sigma_j} C_j(R) + \frac{\sigma_j}{\sigma_k} C_k(R) \right] \nonumber \\ &\times \exp \left( - \frac{z_j - z_k}{\xi_\perp} \right).
\end{align}
Here, $C_j(R)$ and $C_k(R)$ describe the auto-correlation functions of the interfaces $j$ and $k$, while $z_j$ and $z_k$ represent their respective vertical positions. The term $\xi_\perp$ is the vertical correlation length, governing the decay of correlations with increasing vertical distance~\cite{PhysRevB.51.2311}.

\subsection{GID angle calculation}
The (111) plane is parallel to the sample surface due to the texture. The (200) plane forms an angle of $\gamma = 54.7^\circ$ with the (111) plane.
The reciprocal lattice vectors $G_{200}$ will therefore lie on a cone where the angle between the $z$-axis and $G_{200}$ is $\gamma$. If we define the x-axis to line along the incident x-ray beam, the unit vector $\hat{G}_{200}$ can be expressed as 
\[
\hat{G}_{200} = \left( \cos(\phi) \sin(\gamma), \sin(\phi) \sin(\gamma), \cos(\gamma)\right),
\]
where $\phi$ is the angle in the $x-y$ plane measured from the $x$-axis.
The (200) reflection will appear if it simultaneously satisfies the Bragg condition, with $2\theta_{200} = 39.5^\circ$. The $G_{200}$ vectors that would form the Scherrer ring for powder samples lie in the $y-z$ plane and can be expressed as
\[
\hat{G}_{200} = \left( -\sin(\theta_{200}), \sin(\psi) \cos(\theta_{200}), \cos(\psi) \cos(\theta_{200}) \right),
\]
where $\psi$ is the angle in the $y-z$ plane counting from the z-axis.
By solving the above two equations, we find $\phi = \cos^{-1}(\revision{-}\sin(\theta_{200})/\sin(\gamma)) = 114^\circ$ and $\psi = \cos^{-1}(\cos(\gamma)/\cos(\theta_{200})) = 52.2^\circ$.
Using these results, one can calculate the directional unit vector for the outgoing beam $\boldsymbol{k_{out}}$:
\[
k_x = \cos(2\theta_{200}), \quad
k_y = \sin(2\theta_{200}) \sin(\psi), \quad
k_z = \sin(2\theta_{200}) \cos(\psi).
\]
This leads to the angle in the $x-z$ plane (measured from the x-ray axis):
\[
\tan^{-1}(k_z/k_x) = 26.8^\circ,
\]
and the polar angle (measured from the $x-z$ plane):
\[
\tan^{-1}(k_y/\sqrt{(k_x^2 + k_z^2)}) = 30.1^\circ.
\]
We placed the (200) GID detector along this position. 

\subsection{Temperature calculations}
To simulate the temperature evolution in the gold film, we employ a two-temperature model (TTM) coupled with an adaptive Drude-Lorentz model for the dielectric function \cite{Ndione2024}.
The temperature-dependent dielectric function enables the calculation of reflectivity and absorption coefficients as functions of both electron and phonon temperatures.
The captured high-temperature effects are particularly significant at photon energies below the d-band transition threshold in gold, where reduced state blocking leads to a marked decrease in reflectivity and an increase in the absorption coefficient.
The \SI{800}{\nano\meter} wavelength laser pulse is modeled as Gaussian in time with a full width at half maximum (FWHM) of \SI{100}{\femto\second} and an incident fluence of \SI{1.4}{\joule\per\centi\meter\squared}.
For a gold film thickness of \SI{56}{\nano\meter}, homogeneous energy deposition is assumed. 
The temperature-dependent electron heat capacity and electron-phonon coupling parameter are taken from Ref.~\cite{Lin_Zhigilei08}, while the lattice specific heat capacity is assumed constant at \SI{129}{\joule\per\kilogram\per\kelvin}.
The peak electron temperature during the laser irradiation exceeds \SI{15000}{\kelvin}, at which point the heat capacities of the electron and phonon subsystems become comparable.
After \SI{40}{\pico\second}, the material temperature converges to approximately \SI{6600}{\kelvin}, as the electron and lattice temperatures differ by less than \SI{100}{\kelvin}.

\subsection{Elastic coupling between in-plane and out-of-plane strain}

For a linear isotropic elastic solid, the stress–strain relation (Hooke’s law) reads
\begin{equation}
\epsilon_i = \frac{1}{E}\left(\sigma_i - \nu(\sigma_j + \sigma_k)\right),
\end{equation}
where $E$ is Young's modulus, $\nu$ is Poisson's ratio, and $i,j,k$ denote mutually orthogonal axes~\cite{Nix89}.
Here, $\parallel$ and $\perp$ refer to the surface-parallel and surface-normal directions, respectively; tensile strain is taken as positive. 

For a thin film under plane stress conditions, the surface-normal stress vanishes ($\sigma_\perp = 0$) and the two surface-parallel stresses are equal ($\sigma_x = \sigma_y = \sigma_\parallel$). The corresponding strains are
\begin{align}
\epsilon_\parallel &= \frac{1-\nu}{E}\,\sigma_\parallel, \label{eq:strain_parallel}\\[6pt]
\epsilon_\perp     &= -\,\frac{2\nu}{E}\,\sigma_\parallel. \label{eq:strain_perp}
\end{align}
Eliminating $\sigma_\parallel$ gives a direct relation between the surface-parallel and surface-normal strains:
\begin{equation}
\epsilon_\perp = -\,\frac{2\nu}{1-\nu}\,\epsilon_\parallel.
\label{eq:biaxial_strain_relation}
\end{equation}

Using a bulk Poisson’s ratio for gold of $\nu = 0.42$ yields
\begin{equation}
\epsilon_\perp \approx -1.45\,\epsilon_\parallel 
\end{equation}
That is, a longitudinal compression of about $0.5\%$ corresponds to biaxial surface-parallel tensile strain of only $0.35\%$, which corresponds to \(\epsilon_{\mathrm{proj}\_200}(\gamma = \SI{54.7}{\degree}) = 0.5\,\cos^{2}\gamma + 0.35\,\sin^{2}\gamma \approx 0.4\% \). Faurie et al.~\cite{Faurie06} reported significantly higher \textit{effective} Poisson's ratio ($\nu \approx 0.57$) for strongly \{111\}-textured Au films. In that case, $ \epsilon_\perp \approx -2.65\,\epsilon_\parallel$, so $0.5\%$ longitudinal compression would correspond to an even smaller in-plane tensile train of  \(0.19\%\) ( \( \epsilon_{\mathrm{proj}\_200} \approx 0.29\% \) ) .

\begin{acknowledgements}
We acknowledge the European XFEL in Schenefeld Germany, for provision of X-ray free-electron laser beamtime at the Scientific Instrument HED (High Energy Density Science) under proposal number 3082 and would like to thank the staff for their assistance. The authors are grateful to the HIBEF user consortium for the provision of instrumentation and staff that enabled this experiment. 
\revision{Authors acknowledge fruitful discussions with Vanina Recoules.}
\end{acknowledgements}

\begin{funding}
\revision{L.R. acknowledges funding by the German Federal Ministry of Research, Technology and Space (BMFTR) Project No. 05K24PSA.}
D.K. and C.G. acknowledge funding by the Deutsche Forschungsgemeinschaft (DFG) Project No. GU 535/9-1 and No. KS 62/3-1. 
M.B., C.G. and M.N. acknowledge support from DFG Project GU 535/6-1. 
G.J. and M.K. acknowledge funding from DFG Project No. 268565370 (SFB TRR173 Projects A01 and B02) by TopDyn and the BMBF ForLab MagSens. 
T.H., P.D.N., B.R. und S.T.W. acknowledge the support from DFG Project No. 268565370 (SFB TRR173 Projects A08, B03 and INF) and the Allianz für Hochleistungsrechnen Rheinland-Pfalz for providing computing resources through project STREMON on the Elwetritsch high-performance computing cluster.
Ö.Ö. and C.G. acknowledge financial support by the consortium DAPHNE4NFDI in association with the German National Research Data Infrastructure (NFDI) e.V. - project number 4602487. This work was partially supported by the National Research Foundation of Korea (No. RS-2022-00207260, RS-2023-00218180, RS-2025-00516264, RS-2025-02318077).
\\ 
This research was supported in part through the Maxwell computational resources operated at Deutsches Elektronen-Synchrotron DESY, Hamburg, Germany.
\end{funding}

\ConflictsOfInterest{The authors declare no competing interests.}

\DataAvailability{PThe DOI for the original European XFEL data is: 10.22003/XFEL.EU-DATA-003082-00, and will be publicly available after the embargo period of 3 years. \\
}

\bibliography{apssamp} 

\end{document}